\documentclass{article}
\usepackage{cite}
\usepackage{colortbl}
\usepackage[bottom]{footmisc}
\usepackage{graphicx}
\usepackage{hyperref}

\begin{document}

\title{A Distributed System for SearchOnMath Based on the Microsoft BizSpark Program}

\author{Ricardo~M.~Oliveira\quad Flavio~B.~Gonzaga\thanks{Corresponding author (fbgonzaga@bcc.unifal-mg.edu.br).}\\
\\
Universidade Federal de Alfenas\\
Rua Gabriel Monteiro da Silva, 700\\
37130-001 Alfenas - MG, Brazil\\
\\
Valmir~C.~Barbosa\quad Geraldo~B.~Xex\'{e}o\\
\\
Universidade Federal do Rio de Janeiro\\
Caixa Postal 68511\\
21941-972 Rio de Janeiro - RJ, Brazil}

\date{}

\maketitle

\begin{abstract}
Mathematical information retrieval is a relatively new area, so the first search tools capable of retrieving mathematical formulas began to appear only a few years ago. The proposals made public so far mostly implement searches on internal university databases, small sets of scientific papers, or Wikipedia in English. As such, only modest computing power is required. In this context, SearchOnMath has emerged as a pioneering tool in that it indexes several different databases and is compatible with several mathematical representation languages. Given the significantly greater number of formulas it handles, a distributed system becomes necessary to support it. The present study is based on the Microsoft BizSpark program and has aimed, for $38$ different distributed-system scenarios, to pinpoint the one affording the best response times when searching the SearchOnMath databases for a collection of $120$ formulas.

\end{abstract}

\newpage
\section{Introduction}

Unlike textual information retrieval, for which there exist several techniques already widely studied and disseminated, as well as tools capable of tackling the required tasks while performing quite satisfactorily, the area of Mathematical Information Retrieval (MIR) is still in a much less developed stage. In fact, as summarized in Table~\ref{tab:tools}, only in the past few years have techniques for MIR been introduced, usually focusing on very specific problems related to Wikipedia's mathematical pages and indexing around $500\,000$ formulas.

\begin{table}[t]
	\centering
	\caption{Existing Tools for MIR}
	\label{tab:tools}
	\begin{tabular}{ccc}
		\hline
		Reference & Search Domain & No.\ of Formulas\\
		\hline
		\cite{kohlhase:06} & CONNEXIONS project, & $77\,000$ \\
		& functions.wolfram  & $87\,000$\\
		\cite{asperti:06} & Coq proof assistant & $40\,000$ theorems\\
		\cite{pavankumar:12} & Database created & 829 \\
		& by authors & \\
		\cite{schellenberg:12} & $50$ \LaTeX\ documents & $24\,479$ \\
		\cite{kamali:13} & en.wikipedia.org, & $611\,210$ \\
		& DLMF & $252\,148$ \\
		\cite{hu:13} & en.wikipedia.org & $495\,958$ \\
		\cite{lin:14} & en.wikipedia.org, & $521\,782$ \\
		& CiteSeerX & $9\,482$ \\
		\cite{stalnaker:15} & en.wikipedia.org\footnotemark[1] & $482\,364$\\
		\cite{zanibbi:16} & en.wikipedia.org\footnotemark[2]  & $387\,947$\\
		\hline
	\end{tabular}
\end{table}
\footnotetext[1]{Used MREC (Math REtrieval Collection), a collection with approximately $324\,000$ academic publications, during the develompment phase.}
\footnotetext[2]{Includes some information about index size and response time when applied to an arXiv base with about $60$ million formulas.}

As with most niche-oriented forms of information retrieval, MIR has to contend with problems that are specific to the search for mathematical formulas. One of them is the large overhead caused by the various possible uses for the same symbol~\cite{schubotz:16}. These possibilities constitute an important source of ambiguity in MIR, since completely different formulas can be written using essentially the same symbols~\cite{kamali:13}. Another problem is the fact that usually the formulas available on the Web are represented in languages originally conceived with little or no concern for a formula's semantic aspects.

The search engine SearchOnMath\footnote[3]{\url{http://searchonmath.com/}} is one of the most recent arrivals to the field of MIR. Its first version was released in 2013, and by the end of 2015 it had become a start-up. It soon joined the Microsoft BizSpark program, with a modest but very effective monthly allowance, distributed among five email accounts, to hire computers (at most $20$ processing cores). Until 2016, SearchOnMath was able to perform the search for formulas on four databases, namely, the English version of Wikipedia, Wolfram MathWorld, DLMF, and PlanetMath. In such a scenario, only one computer with the so-called A3 Basic configuration of the Microsoft Azure environment was sufficient. This configuration includes a four-core processor, 7 GiB of RAM, and a 120-GB HD.

In the course of 2016, as SearchOnMath began preparations for expansion, a distributed system was developed and tested on Microsoft Azure with the goal of assessing each of the $38$ possible configurations afforded by our constraints within BizSpark. This investigation was based on a set of $120$ preselected formulas that were to be worked on by SearchOnMath within a domain of almost $2$ million formulas, and aimed at discovering which of the candidate configurations was capable of delivering the best response times. Our results and conclusions are presented in this paper.

Our study contributes to the field of MIR in two different ways. The first of them is more of a confirmation of the path we have selected for SearchOnMath. It is therefore of an immediate nature, with short-term applicability by other entrepreneurs. As companies that develop search engines for mathematical formulas begin to appear, mainly as start-ups, it may be reassuring to know that the Microsoft BizSpark program is a very viable opportunity, since it already supports more than $100\,000$ start-ups worldwide and continues to expand. In this regard, information about the infrastructure and operation of SearchOnMath on Azure can be more widely useful. The second contribution is the performance assessment we carried out itself, including the set of $120$ formulas that we put together in order to measure response time, but which can be used for other purposes as well.

\section{Methodology}

For the present study we considered five databases, all obtained throughout the year 2016. Each of these databases is identified in Table~\ref{tab:DB}, along with the respective number of mathematical formulas extracted from it, disregarding repetitions.

\begin{table}[t]
	\centering
	\caption{Database List}
	\label{tab:DB}
	\begin{tabular}{cc}
		\hline
		Database & Number of Formulas \\
		\hline
		Wikipedia, English version\footnotemark[4] & $590\,417$\\
		Wolfram MathWorld\footnotemark[5] & $79\,677$\\
		DLMF\footnotemark[6] & $33\,219$\\
		PlanetMath\footnotemark[7] & $159\,944$\\
		Socratic\footnotemark[8] & $1\,063\,754$\\
		\hline
	\end{tabular}
\end{table}
\footnotetext[4]{\url{http://en.wikipedia.org/}}
\footnotetext[5]{\url{http://mathworld.wolfram.com/}}
\footnotetext[6]{\url{http://dlmf.nist.gov/}}
\footnotetext[7]{\url{http://planetmath.org/}}
\footnotetext[8]{\url{http://socratic.org/}}

\begin{figure}[t]
	\centering
	\includegraphics[width=10 cm]{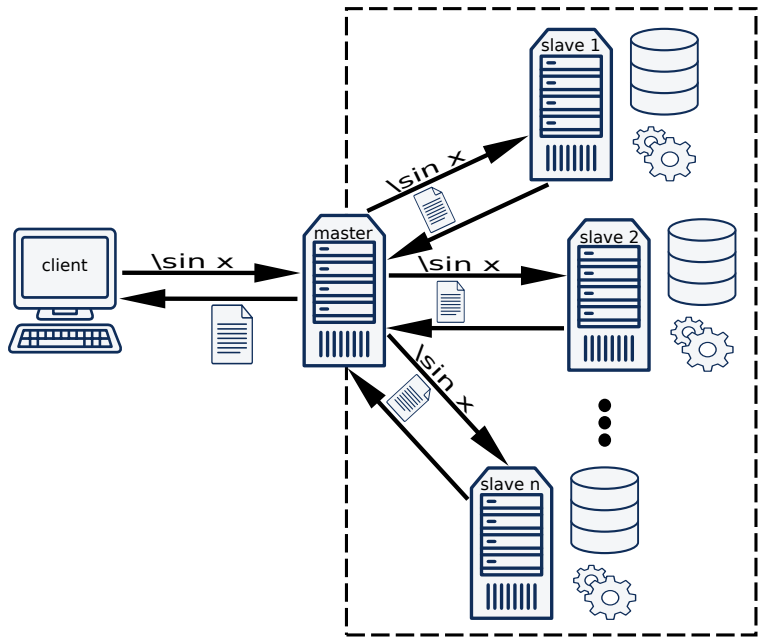}
	\caption{The SearchOnMath architecture.}
	\label{fig:one}
\end{figure}

After the bases were obtained individually, a final database was constructed as the union of all five, still disregarding repetitions. The resulting database contains $1\,905\,358$ indexed formulas. SearchOnMath was then configured as in Figure~\ref{fig:one}. For operation, a client submits a formula to be searched to the master machine, which runs the engine's front-end. After reception by the master, the formula is sent to the slave machines, which do all the necessary processing to find out which formulas in the database are similar to the query formula. The database is distributed across the slaves so that, for example, if we have $10$ of them, then each one has approximately $10\%$ of the formulas. After processing, each slave returns a list containing the most similar formulas it found. The master receives all the lists and then performs the final ordering of the results, returning the consolidated list to the client.

All machines run Linux, and in all cases we configured the master machine with four cores, 7 GiB of RAM, and a 120-GB HD (this configuration is called A3 Basic in Azure). The number of slave machines was obtained based on the remaining allowance resources. We first estimated the amount of our monthly allowance that would correspond to an hour, and then took into account the fact that Azure prices the allocation of machines differently, depending on geographic location. We always chose the region that offered the lowest possible cost in the United States, considering as reference value the one quoted at the date of the beginning of the experiments (Nov.\ 23, 2016). In these circumstances, discounting the A3 Basic cost per hour allowed us the allocation of up to 16 cores to work as slave machines, arranged according to $38$ (out of $65$) different configurations available in the Azure environment.

Table~\ref{tab:configurations} shows all the configurations analyzed for the slave machines. The Configuration column indicates the name of the configuration, its resources detailed in the Cores, RAM, and HD columns. The Machines column indicates the number of machines with this configuration that could be instantiated as slaves. This number is equal to either $\lfloor h/p\rfloor$ or $\lfloor 16/c\rfloor$, whichever is smaller, where $h$ is the available budget per hour, $p$ is the cost per hour of instantiating one machine, and $c$ is the number of cores one machine has.

\begin{table}[p]
	\centering
	\caption{Slave-Machine Configurations and Azure Groups}
	\label{tab:configurations}
	\begin{tabular}{ccccc}
		\hline
		Configuration & Cores & RAM & HD & Machines\\
		\hline
		A0 Basic & 1 & $0.75$ GiB & $20$ GB & 16\\
		A1 Basic & 1 & $1.75$ GiB & $40$ GB & 16\\
		A2 Basic & 2 & $3.50$ GiB & $60$ GB & 8\\
		A3 Basic & 4 & $7.00$ GiB & $120$ GB & 4\\
		A4 Basic & 8 & $14.00$ GiB & $240$ GB & 2\\
		\hline
		A1 v2 & 1 & $2.00$ GiB & $10$ GB & 16\\
		A2 v2 & 2 & $4.00$ GiB & $20$ GB & 8\\
		A4 v2 & 4 & $8.00$ GiB & $40$ GB & 4\\
		A8 v2 & 8 & $16.00$ GiB & $80$ GB & 2\\
		A2m v2 & 2 & $16.00$ GiB & $20$ GB & 8\\
		A4m v2 & 4 & $32.00$ GiB & $40$ GB & 3\\
		A8m v2 & 8 & $64.00$ GiB & $80$ GB & 1\\
		\hline
		D1 v1 & 1 & $3.50$ GiB & $50$ GB & 12\\
		D2 v1 & 2 & $7.00$ GiB & $100$ GB & 6\\
		D3 v1 & 4 & $14.00$ GiB & $200$ GB & 3\\
		D4 v1 & 8 & $28.00$ GiB & $400$ GB & 1\\
		\hline
		D1 v2$^*$ & 1 & $3.50$ GiB & $50$ GB & 14\\
		D2 v2$^*$ & 2 & $7.00$ GiB & $100$ GB & 7\\
		D3 v2$^*$ & 4 & $14.00$ GiB & $200$ GB & 3\\
		D4 v2$^*$ & 8 & $28.00$ GiB & $400$ GB & 1\\
		\hline
		A0 Standard & 1 & $0.75$ GiB & $20$ GB & 16\\
		A1 Standard & 1 & $1.75$ GiB & $70$ GB & 16\\
		A2 Standard & 2 & $3.50$ GiB & $135$ GB & 8\\
		A3 Standard & 4 & $7.00$ GiB & $285$ GB & 4\\
		A4 Standard & 8 & $14.00$ GiB & $605$ GB & 2\\
		A5 Standard & 2 & $14.00$ GiB & $135$ GB & 3\\
		A6 Standard & 4 & $28.00$ GiB & $285$ GB & 1\\
		\hline
		\rowcolor[gray]{0.9}F1$^*$ & 1 & $2.00$ GiB & $16$ GB & 16\\
		\rowcolor[gray]{0.9}F2$^*$ & 2 & $4.00$ GiB & $32$ GB & 8\\
		\rowcolor[gray]{0.9}F4$^*$ & 4 & $8.00$ GiB & $64$ GB & 4\\
		\rowcolor[gray]{0.9}F8$^*$ & 8 & $16.00$ GiB & $128$ GB & 2\\
		\hline
		\rowcolor[gray]{0.6}D11 v1 & 2 & $14.00$ GiB & $100$ GB & 4\\
		\rowcolor[gray]{0.6}D12 v1 & 4 & $28.00$ GiB & $200$ GB & 2\\
		\rowcolor[gray]{0.6}D13 v1 & 8 & $56.00$ GiB & $400$ GB & 1\\
		\hline
		\rowcolor[gray]{0.6}D11 v2$^*$ & 2 & $14.00$ GiB & $100$ GB & 5\\
		\rowcolor[gray]{0.6}D12 v2$^*$ & 4 & $28.00$ GiB & $200$ GB & 2\\
		\rowcolor[gray]{0.6}D13 v2$^*$ & 8 & $56.00$ GiB & $400$ GB & 1\\
		\hline
		\rowcolor[gray]{0.6}G1$^*$ & 2 & $28.00$ GiB & $384$ GB & 1\\
		\hline
	\end{tabular}
\end{table}

Azure groups similar machine configurations~\cite{azure:17}. In Table~\ref{tab:configurations}, a white backdrop indicates machines classified as ``General Purpose---Balanced CPU to memory ratio.'' A light-gray backdrop indicates ``Compute Optimized---High CPU to memory ratio'' machines. Those on a dark-gray backdrop, finally, are ``Memory Optimized---High memory to core ratio'' machines. Configurations with an asterisk (*) by their denominations comprise machines that Azure offers with or without an SSD. Thus, for these machines, both possibilities were evaluated.

\section{Results}
\label{sec:results}

All tests were executed on a set of $120$ formulas\footnote[9]{\url{http://searchonmath.com/formulas}, accessed: Feb.\ 18, 2017.} from~\cite{salem:92, stewart:12, kamali:13, stalnaker:15, pavankumar:12}.

The overall testing scheme for each line of Table~\ref{tab:configurations} (each configuration of the slave machines in Figure~\ref{fig:one}) was the following. The first of the $120$ formulas was submitted for search to the master machine (of type A3 Basic), which then passed it on to the slave machines (of types dependent upon the configuration in question, as per Table~\ref{tab:configurations}) and awaited their results. Having received these, the master machine put together and sorted the final list of results and proceeded to submitting the second formula in the set. This was repeated until all $120$ formulas were searched.

\begin{figure}[p]
	\centering
	\includegraphics[width=10 cm]{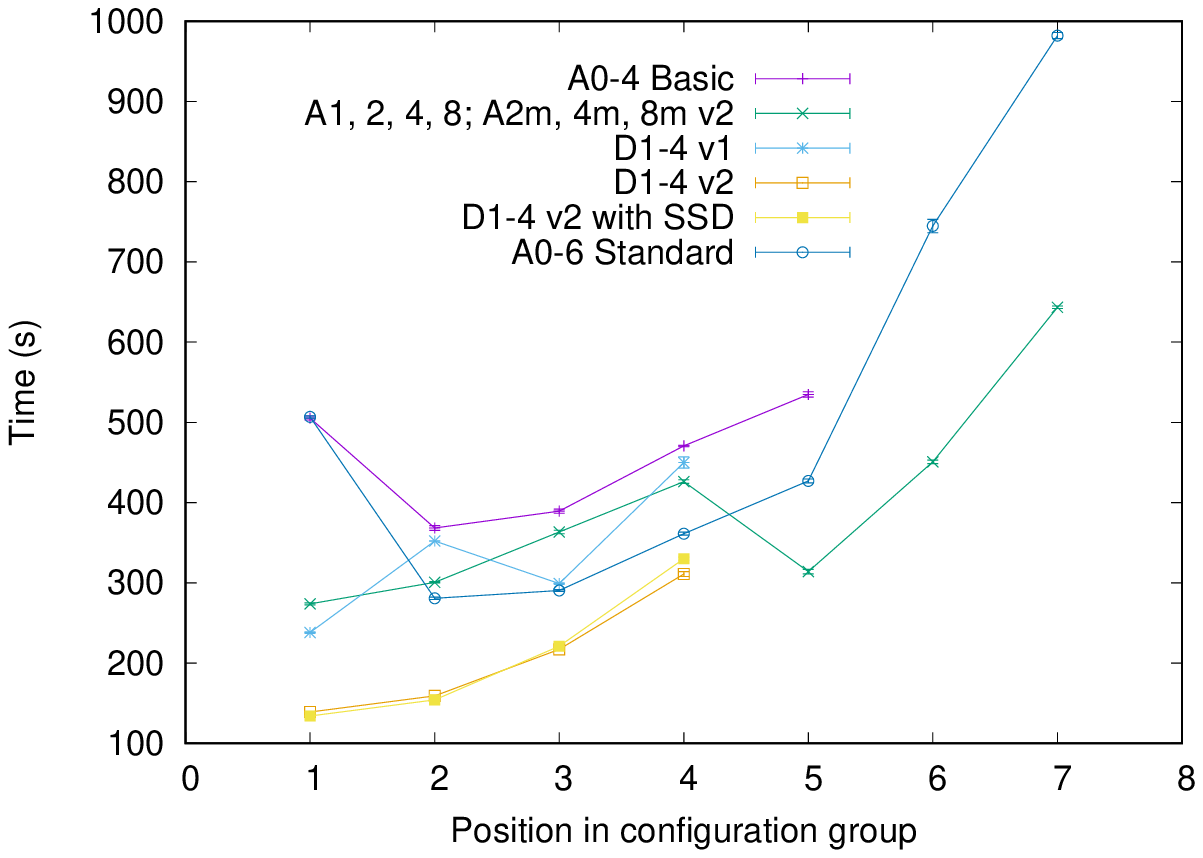}
	\caption{Time spent on General Purpose machines.}
	\label{fig:general}
\end{figure}

\begin{figure}[p]
	\centering
	\includegraphics[width=10 cm]{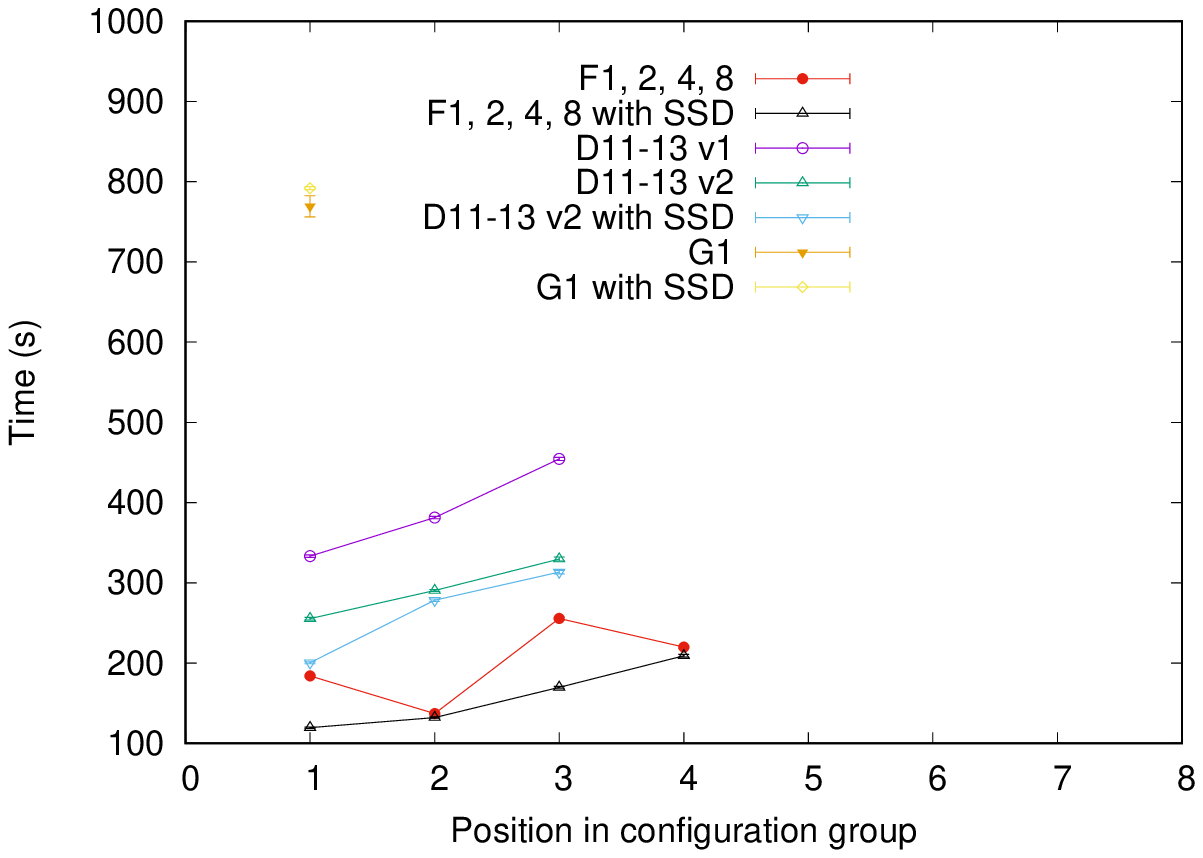}
	\caption{Time spent on Compute Optimized and Memory Optimized machines.}
	\label{fig:memoryOpt}
\end{figure}

This full search pass over all $120$ formulas was repeated $41$ times for each of the configurations of Table~\ref{tab:configurations}. The time spent on each pass was recorded and, at the end, the average time of all $41$ executions was found and its confidence interval estimated (at the $99\%$ level). We note that each time measurement disregards every communication delay between the client and the master (cf.\ Figure~\ref{fig:one}). As a result, all time figures we report are search-related, referring to processing time at the master or at a slave, or to internal network delays of the distributed system. We give results in Figures~\ref{fig:general} and~\ref{fig:memoryOpt}, respectively for the machines of Azure type General Purpose and for those of the other two types (Compute Optimized and Memory Optimized).

Each plot in these figures refers to a group of slave-machine configurations, as implied by the horizontal rules in Table~\ref{tab:configurations}, and positions each of the group's configurations on the abscissa axis in the order given in the table. So, for example, configurations A0--4 Basic are grouped together, with A0 Basic appearing leftmost in Figure~\ref{fig:general}, followed by A1 Basic, and so on. It is also worth noting that the confidence intervals are often negligible and therefore hard to discern in the figures.  

As it turns out, the best scenario for the SearchOnMath system is the slave-machine configuration F1 with SSD, which comprises $16$ identical single-core machines, each with 2 GiB of RAM and a 16-GB SSD. With this configuration, the time needed to search for the $120$ formulas was about $120$ seconds on average (roughly $1$ second per formula), with a confidence interval of approximately $\pm 0.86$ seconds.

Notwithstanding this, we note that in general the SSD-based configurations did not result in a large difference when compared to their HD-based counterparts. This was expected, given that SearchOnMath carries the formulas in memory while running, thus considerably reducing the need for access to secondary storage. Two exceptions to this note occurred for configurations F1 and F4, in which case time differences were indeed significant. Nevertheless, we are unable to explain such differences on grounds of the SearchOnMath algorithms, and must therefore speculate that they have to do with factors internal to Azure.

\section{Conclusions}

Carrying out the experiments described in this paper has allowed us to observe the functioning of SearchOnMath on a variety of configurations of the Microsoft Azure cloud environment. We experimented with all configurations compatible with our BizSpark status and, within these limits, identified a configuration capable of supporting $1$-second searches for $120$ (out of just over $1\,900\,000$) formulas. At the relatively modest cost currently afforded us by the Microsoft BizSpark program, these experiments will help us envisage plans to scale up operations. We note, finally, that the $120$ formulas selected for the experiments will remain available from SearchOnMath for possible future use in further comparative studies.

\subsection*{Acknowledgments}

We thank the Federal University of Alfenas and NidusTec Business Incubator, as well as CNPq, CAPES, and a FAPERJ BBP grant for financial support. We also thank Microsoft for the opportunity to participate in their BizSpark program.

\bibliography{searchonmath}
\bibliographystyle{unsrt}

\end{document}